# AI-enhanced High Resolution Functional Imaging Reveals Trap States and Charge Carrier Recombination Pathways in Perovskite

*Qi Shi[1,2]\*, and Tönu Pullerits [1,2]\**

[1] The Division of Chemical Physics, Department of Chemistry, Lund University, Kemicentrum Naturvetarevägen 16, 223 62 Lund, Sweden

[2] NanoLund, Lund University, Professorsgatan 1, 223 63 Lund, Sweden

AUTHOR INFORMATION

**Corresponding Author**

\*qi.shi@chemphys.lu.se

\*Tonu.Pullerits@chemphys.lu.se

**Abstract**

Understanding and managing charge carrier recombination dynamics is crucial for optimizing the performance of metal halide perovskite optoelectronic devices. In this work, we introduce a machine learning-assisted intensity-modulated two-photon photoluminescence microscopy (ml-IM2PM) approach for quantitatively mapping recombination processes in $MAPbBr_3$ perovskite microcrystalline films at micrometer-scale resolution. To enhance model accuracy, a balanced classification sampling strategy was applied during the machine learning optimization stage. The trained regression chain model accurately predicts key physical parameters—exciton generation rate ($G$), initial trap concentration ($N_{TR}$), and trap energy barrier ($E_a$)—across a 576-pixel spatial mapping.

These parameters were then used to solve a system of coupled ordinary differential equations, yielding spatially resolved simulations of carrier populations and recombination behaviours at steady-state photoexcitation. The resulting maps reveal pronounced local variations in exciton, electron, hole, and trap populations, as well as photoluminescence and nonradiative losses. Correlation analysis identifies three distinct recombination regimes: (i) a trap-filling regime predominated by nonradiative recombination, (ii) a crossover regime, and (iii) a band-filling regime with significantly enhanced radiative efficiency. A critical trap density threshold (~$10^{17}$ cm³) marks the transition between these regimes. This work demonstrates ml-IM2PM as a powerful framework for diagnosing carrier dynamics and guiding defect passivation strategies in perovskite materials.

# 1 Introduction

Since 2012[1], the power conversion efficiency of perovskite photovoltaic (PV) solar cells and the external quantum efficiency of perovskite light-emitting devices (LEDs) have experienced a remarkable increase, reaching 26.1%[2] and 28.9%[3], respectively. The rapid developments of perovskite technology, together with the cost-effective solution-based material synthesis methods, is an important direction to unlock the full potential of many sustainable future technologies[4]. However, a challenge with such facile solution-based synthesis methods and the practical utilization of perovskite material is that defects are unavoidably formed in the material. The malignant defects that trap energized electrons and cause nonradiative recombination (NRR) losses– where electron energy is dissipated as heat instead of being emitted as photons– exist in the perovskite-based devices thus limiting the device efficiency[5]. A thorough understanding of the defect properties and their impact on performance of the materials is the key to optimization and future success of such technologies[6–8].

Probing defects in a perovskite semiconductor can be achieved through exciting electrons in the material by light absorption and monitoring photoluminescence (PL) as the electrons radiatively return to their ground state[7,9,10]. An important dimension here is sufficiently high spatial resolution since the wet chemistry methods lead to morphological inhomogeneities[11–14]. This means that different parts of the material may have very different properties. For example, the subpopulation of dark grains and boundaries during the perovskite growth work as nonradiative recombination pathways which limit the PL emission[12]. Besides, the pronounced spatial variations in the photocurrent response due to differences in the perovskite grains at a sub-micrometer scale have been demonstrated[13]. In addition, spatial variations of free charge carrier and exciton populations within perovskite grains at a sub-micrometer scale have been verified as well[15]. Such effects would be averaged out when using traditional ensemble spectroscopy methods. Consequently, it is imperative to develop and apply methods that enable the probing of the defect properties and differentiation of defects' influence on the PL intensity of these materials with sufficient spatial resolution.

Newly developed intensity modulation two-photon excited PL microscopy (IM2PM) has been applied to the perovskite films to study the recombination processes[13,15,16]. Since the two-photon excitation has a micrometer resolution and the measured PL is directly related to the functioning of the studied material, we call the technique functional mapping. The excitation modulation leads to oscillating PL signals which are analysed in terms of Fourier components containing information about photo-induced charge carrier trapping and recombination processes. Provided the materials and measured signals, physical models are formulated to describe such dynamics. The models link the IM2PM experiment with detailed material properties. Because of the complexity of the models, the usual direct data fitting algorithms are of limited applicability here. Machine learning (ML) methods have recently gained prominence playing an important role in bridging the gap between complex experimental data and a deeper understanding of the dynamic processes in the studied materials[17–20]

Significant advancements have been made in the field of AI-enhanced functional microscopy, including the successful application of the machine learning regression analyses for intensity modulation two-photon microscopy (ml-IM2PM). This AI enhanced IM2PM approach utilizes a robust model to capture carrier trapping and charge carrier recombination processes in the $MAPbBr_3$ perovskite microcrystalline film, complemented by a regression chain that incorporates extra tree regression analysis.[18] The defect concentration as well as the defect energy barrier parameters are predicted by the ml-IM2PM model. Thereby, ml-IM2PM provides valuable insights into details of trapping, detrapping, and nonradiative recombination processes, offering a comprehensive understanding of perovskite material photophysics.

In this work, we apply ml-IM2PM to investigate micrometer-scale charge carrier dynamics in MAPbBr$_3$ perovskite polycrystalline films. To enhance the accuracy and generalizability of the model, a balanced classification sampling strategy is employed during the machine learning optimization process, ensuring more uniform coverage across a broad range of exciton generation rates. By spatially mapping key recombination parameters, including exciton generation rate, initial trap concentration, and trapping energy barrier—we reveal the intrinsic heterogeneity in local recombination behavior across the film. Using a time-dependent physical model, we quantify the populations of photogenerated carriers at photoexcitation equilibrium at room temperature, such as electrons, holes, excitons, and filled traps, and evaluate their roles in radiative and nonradiative recombination. Correlation analysis of the spatially resolved data reveals two dominant recombination regimes: a trap-predominated regime, where nonradiative losses are predominant, and a band-filling regime, characterized by enhanced radiative PL emission. We further identify a threshold in trap density that marks the transition between these regimes. These findings provide a quantitative framework for understanding how local defect landscapes influence recombination dynamics and offer valuable insight for guiding defect passivation and performance optimization strategies in perovskite optoelectronic devices.

## 2 Methods.

### 2.1 Perovskite

#### 2.1.1 Temperature dependent variation on trap density

The performance of LHP devices is significantly affected by the type and density of defects in the material. Defects that trap charge carriers, especially those that form deep energy levels, can lower the efficiency of devices. In solution-based synthesis of perovskites, the number of defects depends on how easily they form (defect formation energies) and the conditions during crystal growth. Defects with lower formation energies are more likely to appear and will be present in higher concentrations.

For MAPbBr$_3$ perovskite with 1:1 ratio of lead bromide (PbBr$_2$) and methylammonium bromide (MABr)) precursors, studies have shown that defects such as MA and bromide interstitials, $MA_i$ and $Br_i$, (where extra MA or bromide ions are located in positions they don't normally occupy) show lower formation energies and appear more frequently [21,22]. On the other hand, defects like lead and bromide vacancies (missing Pb or Br atoms), $V_{Pb}$ and $V_{Br}$, have higher formation energies and therefore appear less frequently. Other possible defects, such as lead interstitials (extra Pb atoms) and methylammonium vacancies (missing MA ions), $Pb_i$ and $V_{MA}$, are even less likely because they require more energy to form.

Due to the inherently soft lattice structure of metal halide perovskites [23], these materials exhibit a dynamic defect picture where trap density is not fixed. Instead, defects may be passivated or activated by factors such as ion/defect migration, light exposure, or interaction with molecular oxygen[21,24–27]. Besides, trap density decreases at lower temperature has been revealed by several studies. [21,28–32] It has been suggested that lowering the temperature tends to "freeze" ions and suppress ion migration (e.g., I$^-$ or Pb$^{2+}$)[33], leading to the passivation of point defects. Additionally, carrier trapping at iodine interstitials is influenced by the trapping kinetic barrier, which involves overcoming a small energy threshold of about 0.10 eV to reach a metastable neutral iodine state ($I_i^0$) where the electron is trapped[29,30]. At lower temperatures, the reduced thermal energy limits iodine's ability to capture mobile carriers, effectively decreasing the trap density. Consequently, the temperature-dependent effective trap density can be approximated by a Boltzmann equilibrium:

$$N_{Tr} = N_{TR} \times e^{-\frac{E_a}{k_B T}}.$$

where, $E_a$ represents either the trapping kinetic barrier or the ion migration barrier, and we call it trap energy barrier, $N_{TR}$ is the initial trap density after sample synthesis, and $N_{Tr}$ is the total trap density at experimental temperatures ($T$).

## 2.1.2 Charge carrier recombination model

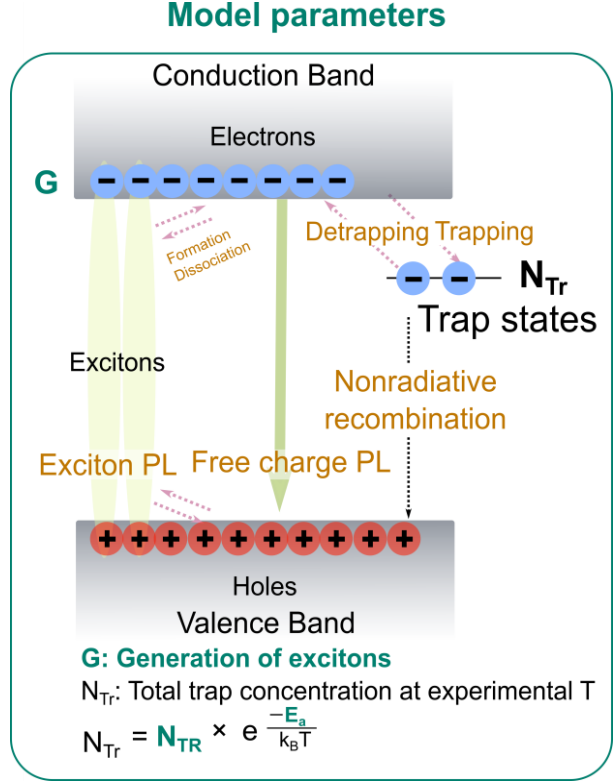

**Figure 1.** The charge carrier recombination model for MAPbBr$_3$ perovskite microcrystalline films. The following model parameters are obtained from analyses: generation rate of excitons (G), initial trap density after sample synthesis ($N_{TR}$), and defect energy barrier ($E_a$). The model also includes recombination processes, including exciton PL emission (PL-ex), free electron-hole PL emission (PL-eh), trapping, detrapping, and non-radiative recombination (NRR) loss. The parameters determining those processes are obtained from literature.

Figure 1 describes a charge carrier recombination model that captures the key features of photo-induced dynamics in MAPbBr$_3$ perovskite microcrystalline films.[18],[7] The following set of coupled ordinary differential equations (ODEs) describes the dynamics of the photogenerated excitons $n_{ex}$, free electrons $n_e$ and filled traps $n_{Tr}$:

$$\frac{dn_{ex}}{dt} = G + R_f n_e n_h - R_d n_{ex} - R_{ex} n_{ex}, \tag{1}$$

$$\frac{dn_e}{dt} = R_d n_{ex} - R_f n_e n_h - \gamma_{trap} n_e (N_{Tr} - n_{Tr}) + \gamma_{detrap} n_{Tr} - R_{eh} n_e n_h, \tag{2}$$

$$\frac{dn_{Tr}}{dt} = \gamma_{trap} n_e (N_{Tr} - n_{Tr}) - \gamma_{detrap} n_{Tr} - \gamma_{NRR} n_{Tr} n_h. \tag{3}$$

The parameters $R_f$, $R_d$, $R_{eh}$, $R_{ex}$, $\gamma_{trap}$, $\gamma_{detrap}$ and $\gamma_{NRR}$ are explained and described in Table 1. The values of these parameters are based on the available literature data. Since there is quite some spread of the reported parameter values in the literature, the concrete numbers are to some degree uncertain. This variability arises from differences in experimental conditions, material properties, and measurement techniques, which can influence the parameter estimates [34–36]. For example, electron hole radiative recombination rate $R_{eh}$ for MAPbBr$_3$ perovskite materials exhibits significant variation with different thicknesses for thin film and different grain sizes for polycrystalline sample: $4.9 \pm 0.2 \times 10^{-10}$ [36] (for ~ $86 \pm 11$ nm thickness film), $2.6 \times 10^{-10}$ (for ~ 538 nm thickness film) [35], or $1.0 \pm 0.9 \times 10^{-10}$ (for ~1 μm crystal size), $3.7 \pm 0.2 \times 10^{-11}$ (for ~$50 \pm 25$ nm crystal size), and $1.3 \pm 0.1 \times 10^{-10}$ (for ~$9.1 \pm 1.6$ nm crystal size) [34] $cm^3 s^{-1}$, respectively.

Three model parameters are not known to sufficient extent and will be determined by ml-IM2PM. The first parameter, G, represents the generation rate of excitons, which is determined by both experimental condition and material properties. The irregularities in the micrometer-scale surface morphology of a perovskite microcrystal film may significantly affect two-photon absorption and consequently G. In addition, the total defect concentration $N_{TR}$ and the defect energy barrier $E_a$ are not known and will be determined.

Our analysis specifically focuses on electron trapping [8,37,38]. Electron trapping leads to an imbalance in carrier concentration. Each trapped electron leads to a corresponding free hole in the valence band. As a result, the concentration of free holes is larger than the concentration of free electrons by an amount equal to the concentration of trapped electrons: $n_h = n_e + n_{Tr}$. The consequent accumulation of long-lived trapped charge carriers leads to the increased hole concentration, a phenomenon known as the photodoping, and can significantly enhance PL emission.[39–41] Hole traps are not considered, but their influence under similar conditions would be similar to that of electron traps. The PL emission originates from both exciton and free electron-hole pair radiative recombination (PL ~$R_{ex} n_{ex} + R_{eh} n_e n_h$) [42,43]. Auger recombination is not included in the model due to the low concentration of two-photon excitation.

Table 1: Summary of the parameters for MAPbBr$_3$ hybrid metal halide perovskite ml-IM2PM analysis. *RT* refers to room temperature.

| Recombination Parameters | Values | Refs |
|---|---|---|
| Exciton formation rate $R_f$ | $10^{-12}\ cm^3 s^{-1}$ | [7] |
| Exciton dissociation rate $R_d(RT)$ | $5 * 10^{11}\ s^{-1}$ | [44] |
| Trapping rate $\gamma_{trap}$ | $8 * 10^{-9}\ cm^3 s^{-1}$ | [8] |
| Detrapping rate $\gamma_{detrap}(RT)$ | $10^7\ s^{-1}$ | [8] |
| Electron hole radiative recombination rate $R_{eh}$ | $5 * 10^{-11}\ cm^3 s^{-1}$ | [8] |
| Exciton radiative recombination rate $R_{ex}$ | $5 * 10^7\ s^{-1}$ | [10] |
| Shockley-Read-Hall (SRH) nonradiative recombination rate $\gamma_{NRR}(RT)$ | $5 * 10^{-9}\ cm^3 s^{-1}$ | [8] |
| Exciton binding energy | 84 meV | [45] |

The model assumes the exciton dissociation-association equilibrium, which follows the detailed balance relation of the Saha-Langmuir (S-L) equilibrium model[46,47] expressed as:

$$\frac{R_d(T)}{R_f} = C * e^{-\left(\frac{E_b}{kT}\right)}. \tag{4}$$

Here $E_b$ is the exciton binding energy 84 meV [45]. We use the known values of $R_d(RT)$ and $R_f$ to calculate the parameter C at room temperature (RT). Once C is determined, the temperature-dependent $R_d(T)$ can be calculated accordingly.

**2.2 IM2PM**

IM2PM is an advanced imaging technique that combines two-photon excitation with intensity modulation to study photo-induced dynamics. It has been used to investigate photoactive materials with complex photophysical properties, like hybrid halide perovskite[13,15,16,18]. In IM2PM, the intensity of a pulsed laser is modulated by an envelope function $I \sim (1 + \cos(2\pi\phi t))$ at a specific frequency $\phi$. This modulation induces an oscillatory exciton generation rate within the sample, following a similar envelope function, $G \sim (\beta I^2)$, where $\beta$ represents the two photon absorption coefficient of MAPbBr$_3$ perovskite, and G is the exciton generation rate at the maximum of laser intensity modulation. Additionally, PL signal from the sample oscillates at harmonic frequencies of the laser intensity modulation. Fourier analyses of the IM2PM reveal details about the kinetics of charge carrier trapping and recombination processes within the material, offering a way to quantify complex, time-dependent population dynamics.

The specifics of the experimental setup have been described elsewhere[15,16,18] and are summarized in Section S1. In our current experiment, the excitation pulses are 14.25 ns apart. At the arrival of a pulse, a significant fraction of the carriers generated by the previous pulses may have remained. In particular the trapped carriers can recombine very slowly. Previous studies have shown that the inclusion of the carrier accumulation effects is important for an accurate description of carrier recombination dynamics [8,16,18,48]. In the IM2PM simulations as well as the ml-IM2PM assisted retrieval processes, the accumulation effects are explicitly taken into account.[18]

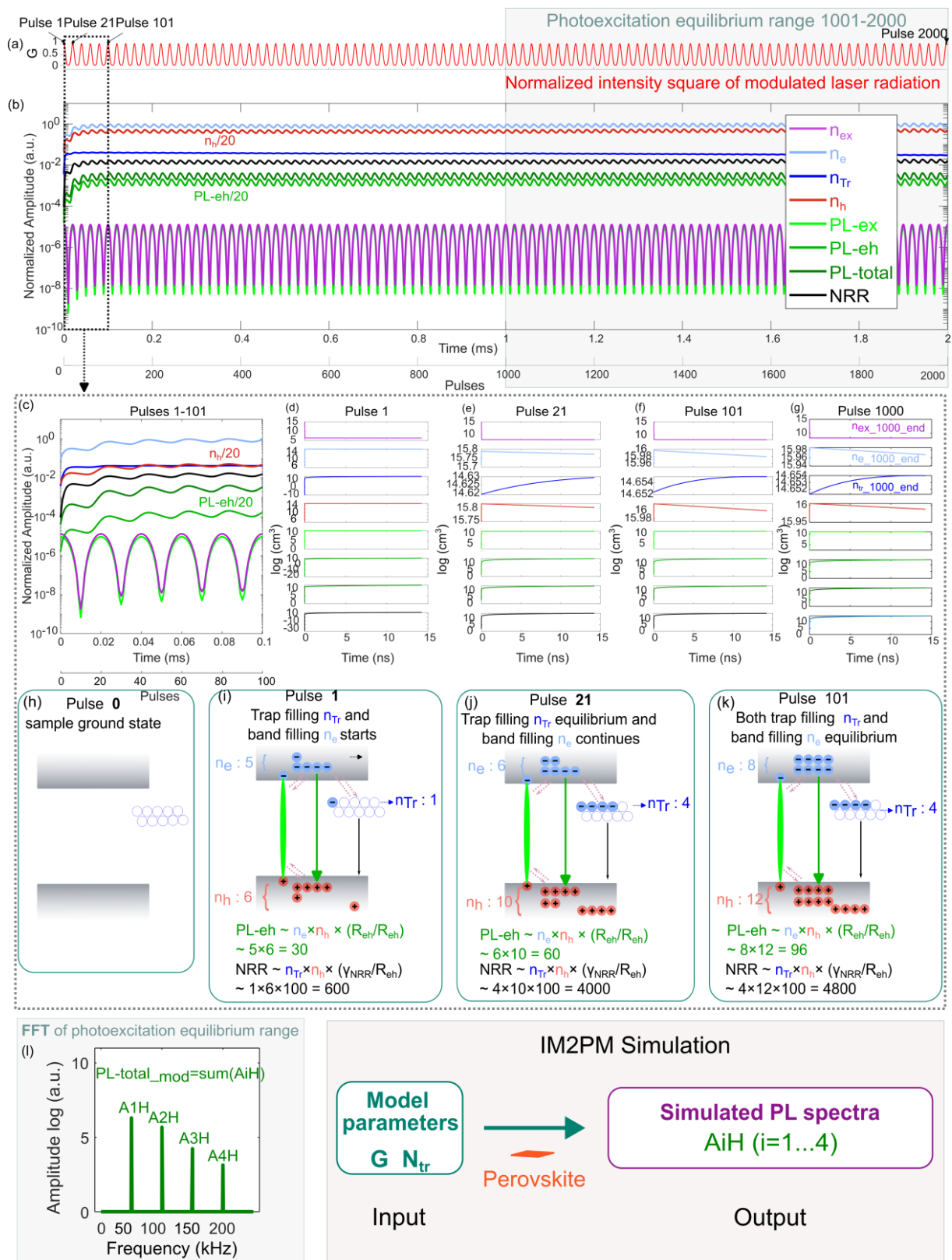

**Figure 2.** Visualization of IM2PM simulations of charge carrier dynamics in MAPbBr$_3$ microcrystals. The charge carrier recombination model parameters (top left panel, marked in green) govern the recombination processes (bottom right panel, marked in orange), with the IM2PM simulation processes shown in panel (a) to (l). The X-axis in panels (a), (b), and (c) represents time in pulse intervals. During the actual experiment, approximately 1400 pulses excite the sample over a single intensity modulation

period (20 microseconds). For clarity, the simulation illustrates only 20 pulses per modulation period in this figure. (a) The red line represents the excitation profile, where the envelope displays the normalized intensity square of the modulated laser radiation, corresponding to the normalized generation rate of excitons at the maximum of the laser intensity modulation (G). (b) The modulated charge carrier dynamics are depicted, including the populations of excitons ($n_{ex}$, pink), electrons ($n_e$, sky blue), holes ($n_h$, light coral), and filled traps ($n_{Tr}$, blue). Additionally, the envelopes of the modulated two-photon excited PL from excitons (PL-ex, light green), PL from electron-hole recombination (PL-eh, lime green), the total PL signal (PL-total, forest green), and nonradiative recombination depopulation loss (NRR, black) are shown. The parameters governing these simulations, as derived from ODEs, are listed in Table S1. (c) A zoomed-in view of pulses 1 to 101 from (b) illustrates the progressive evolution of carrier populations. (d–g) The transient charge carrier dynamics at pulse 1, 21, 101, and 1000 are depicted, showing how carrier populations change over time. The exciton, free electron, and filled trap populations at the end of pulse 1000—denoted as *n_x_1000_end*, *n_e_1000_end*, and *n_Tr_1000_end*, respectively—serve as the initial conditions for the approximate photoexcitation equilibrium regime spanning pulses 1001 to 2000. (h–k) The simplified carrier distributions at pulse 0, 1, 21, and 101 are illustrated, demonstrating the interplay between exciton dissociation, trap filling, and band filling at different time intervals. (l) The fast Fourier transform (FFT) of the photoexcitation equilibrium regime (pulses 1001–2000) reveals that the modulated carrier populations result from the summation of four harmonic signals ($\sum_{i=1}^{4} AiH$). The bottom right panel summarizes the simplified workflow of IM2PM simulation.

Figure 2 illustrates the simulated charge carrier dynamics in MAPbBr$_3$ microcrystals under IM2PM. The excitation profile is shown in Figure 2a, where the red line represents the envelope of the normalized intensity square of the modulated laser radiation, which dictates the generation rate of excitons (G). Due to the periodic intensity modulation of the laser pulses, oscillations appear in carrier populations (excitons ($n_{ex}$, pink), electrons ($n_e$, sky blue), holes ($n_h$, light coral), and filled traps ($n_{Tr}$, blue)$_r$), PL emissions (PL from excitons (PL-ex, light green), PL from electron-hole recombination (PL-eh, lime green), the total PL signal (PL-total, forest green)), and nonradiative recombination depopulation loss (NRR).

To analyse the transient charge carrier dynamics, Figure 2c presents a zoomed-in view of pulses 1 to 101 from Figure 2b, providing a clearer picture of the stepwise changes in carrier populations. Initially, before laser excitation, the system is in the ground state (Figure 2h), where all carrier concentrations are zero. Upon excitation by pulse 1 (Figure 2d), a large number of excitons are generated. However, excitons rapidly dissociate into electron-hole pairs[44], leading to a sharp decrease in the exciton population while simultaneously increasing the electron, hole, and filled trap population. By the end of pulse 1, a significant fraction of electrons and filled traps remain in the system, marking the onset of trap filling and band filling. These remaining electrons contribute to band filling, while remaining filled traps contribute to trap filling. The integration of the eight populations along the interval time between two laser pulses (14.25 ns) results in the first data points in Figures 2b and 2c, representing the first step of the total carrier population evolution. At the end of pulse 1 (Figure 2d), we simplify the carrier distribution by considering 5 remaining electrons in the conduction band and 1 filled trap in the trap states in Figure 2i. The PL-eh emission can be approximated as the product of electron and hole populations and normalized by the radiative recombination rate with a ratio $\frac{R_{eh}}{R_{eh}}$, calculated as $5 \times 6 = 30$. In comparison, the NRR can be estimated and normalized by multiplication the filled trap population, hole population, and the ratio between the nonradiative recombination rate $\gamma_{NRR}$ and the radiative electron-hole recombination rate $R_{eh}$ ($\frac{\gamma_{NRR}}{R_{eh}} = 100$, Table 1), yielding $1 \times 6 \times 100 = 600$.

From Figures 2b–c, the integrated filled trap population reaches equilibrium at approximately pulse 21, while the integrated electron population reaches equilibrium at approximately pulse 101. Comparing the charge carrier dynamics at pulse 21 (Figure 2e) and pulse 101 (Figure 2f), we observe that at pulse 21, the filled trap population stabilizes at approximately $10^{14.6} \approx 4 \times 10^{14}\ cm^3$. Therefore, we simplify the carrier distribution in Figure 2j by considering an increased electron population of 6 and an equilibrated filled trap population of 4. Consequently, the PL-eh and NRR can be approximated as 60 and 4000, respectively. Comparing the charge carrier dynamics at pulse 101 (Figure 2f) and pulse 1000 (Figure 2g), we find that at pulse 101, the electron population reaches equilibrium, approximately $10^{15.97} \approx 9 \times 10^{15}\ cm^3$. Therefore, in Figure 2k, we simplify the carrier distribution by considering an increased and equilibrated electron population of 8 and an equilibrated filled trap population of 4. At this stage, the PL-eh and NRR can be approximated as 96 and 4800, respectively.

At the end of pulses 1000 (Figure 2g), the exciton ($n_x\_1000\_end$), free electron ($n_e\_1000\_end$), and filled trap ($n\_Tr\_1000\_end$) populations are taken as the initial conditions for the subsequent photoexcitation equilibrium regime observed between pulses 1001 and 2000. Within such regime, a FFT analysis is performed to further investigate the frequency response of the system (Figure 2l). The modulated carrier response is decomposed into a constant DC background and four Fourier components (A1H, A2H, A3H, A4H), which correspond to the first four harmonics of the modulation frequency. The total modulated PL emissions from both exciton and electron hole pairs therefore is the summation of four harmonic signals ($\sum_{i=1}^{4} AiH$) of the simulated forest green ossilation in Figure (2b-2c). In summary, with the model parameters from ODEs, the IM2PM PL (AiH) spectra can be simulated, which can compare with the experimental PL (AiH) spectra.

**2.3 ml-IM2PM**

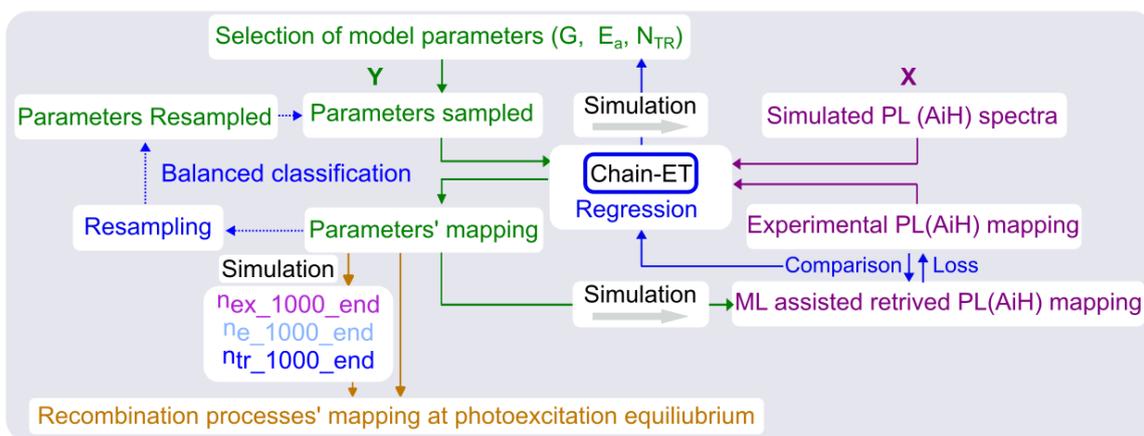

**Figure 3.** Illustration of the ml-IM2PM workflow for MAPbBr₃ perovskite microcrystalline films.

Figure 3 explains the ml-IM2PM [18] workflow applied to MAPbBr₃ perovskite microcrystalline films to extract spatially resolved recombination parameters. The process begins with defining three key model parameters: the exciton generation rate ($G$), the trap energy barrier ($E_a$), and the initial trap density after synthesis ($N_{TR}$). To capture a broad range of material behaviors, these parameters are sampled across defined distributions: five linearly spaced G values, eight linearly spaced $E_a$ values, and seven logarithmically spaced $N_{TR}$ values. Using these combinations, a synthetic dataset of a wide range of temperature-dependent PL Fourier (AiH) spectra is generated through IM2PM simulations (as detailed in Figure 2). This synthetic dataset is used to train a supervised machine learning model, forming the first step of the ml-IM2PM workflow. A Chain-Extra Trees regressors is optimized to predict the selected three model parameters. The validity of the predictions was confirmed by reconstructing temperature-dependent PL (AiH) spectra from the predicted parameters, which closely matched the experimental temperature dependent PL (AiH) measurements (Figure S3), with average Pearson correlation coefficient of 0.928 across eight temperatures, and 0.967 at room temperature.

With the correlation studies of the predicted parameters, uneven sampling of G led to imbalanced coverage in the parameter space (Figure S4). To improve robustness, alternative G sampling strategies were tested and validated through a three-step cross-checking procedure (detailed in SI, Section S4). A logarithmic sampling of G was ultimately selected, as it provided the most balanced distribution and highest agreement with experimental PL (AiH) data, with averaged PCC of 0.950 across eight temperatures, and 0.974 at room temperature (Table S2 & Figure S5). This optimized ml-IM2PM approach enables accurate and spatially resolved mapping of key recombination parameters in complex perovskite films.

Microscale spatial mapping of the model parameters extracted from the ml-IM2PM analysis is presented in Figure 4. These spatially resolved parameters serve as inputs for solving the system of ordinary differential equations (ODEs) that describe charge carrier recombination dynamics. Specifically, the exciton, electron, and filled trap populations at the end of the 1000th excitation pulse—captured during IM2PM simulations—are used as initial conditions for time-dependent simulations. By numerically integrating the ODE system from 0 to 1 millisecond, the spatial evolution of key carrier populations is obtained under photoexcitation equilibrium conditions at room temperature.

The resulting spatial distributions of charge carrier populations and recombination processes are shown in Figure 5. These include the densities of excitons ($n_{ex}$), free electrons ($n_e$), holes ($n_h$), and filled traps ($n_{Tr}$), as well as the total modulated photoluminescence (PL-total) and nonradiative recombination losses (NRR). This spatially resolved modeling provides a quantitative framework for visualizing local charge dynamics and identifying regions of performance-limiting recombination behavior across the film.

## 3 Results and discussion

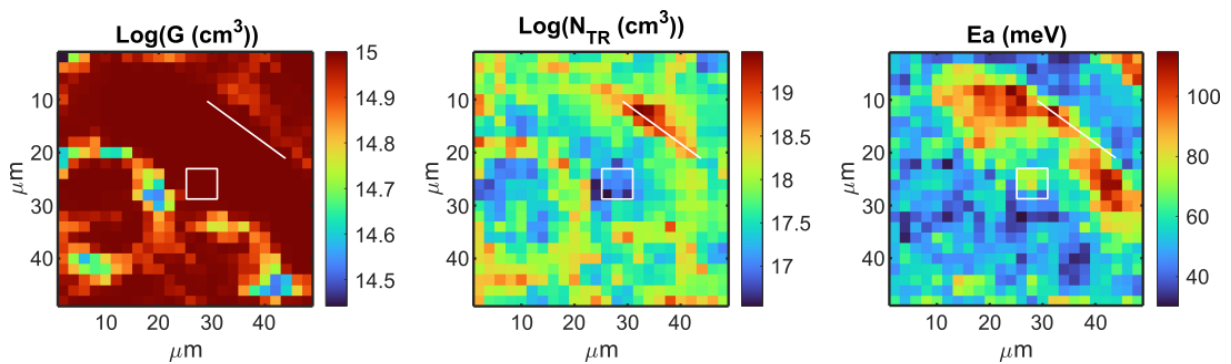

**Figure 4.** Micro-meter scale maps of the model parameters obtained from ml-IM2PM analyses. (a) generation rate of excitons (G), (b) initial trap concentration after sample synthesis ($N_{TR}$), and (c) trap energy barrier ($E_a$).

Figure 4 shows the spatial distribution of exciton generation (a), initial trap concentration after sample synthesis (b) and trap energy barrier (c). The images reflect high morphological heterogeneity which may be related to micrometer-scale morphology irregularities. The microscope used in this study has a depth of focus (DOF) approximately 3.2 μm, given by $\frac{n\lambda}{NA^2}$, where $n$ is the refractive index in the object space (1, air), $\lambda$ is the imaging wavelength (800 nm), $NA$ is the numerical aperture (0.5, Edmund Optics)[49,50]. The DOF is similar to the length scale of the surface irregularities and therefore significantly affects the two-photon absorption, therefore the generation rate of excitonstons (G) in Figure 4a.

The initial trap concentration $N_{TR}$, shown in Figure 4b, ranges from $10^{16}\ cm^{-3}$ to $10^{19}\ cm^{-3}$, which is consistent with trap densities determined by PL emission reported in several studies [51–53]. In the white line region, the initial trap concentration $N_{TR}$ is about $10^{19}\ cm^{-3}$. As discussed in section 2.1.1, $MA_i$ and $Br_i$ defects with lower formation energy could contribute to the higher total trap concentration in the white line region. In the region marked by a white square, the initial trap concentration $N_{TR}$ is about $10^{17}\ cm^{-3}$. Here, defects such as $V_{Pb}$ and $V_{Br}$ with higher formation energy may predominate, therefore the lower trap concentration.

The trap energy barrier $E_a$, shown in Figure 4c, ranges from 30 meV to 110 meV, which aligns well with bromide vacancy-assisted Br migration energy barrier reported in several studies providing values in this range: 31 meV,[54] 73 meV[54] and 90 meV [25]. In the white line region, the energy barrier is about 80 meV and the total trap density $N_{Tr}$ is about $1.3 \times 10^{17}\ cm^{-3}$. While in the line square region, the energy barrier is about 60 meV[54][54] and the total trap density $N_{Tr}$ is about $7 \times 10^{15}\ cm^{-3}$. Differences in the reported ion migration energy barriers across studies most likely originate from the variations in fabrication methods and measurement conditions.

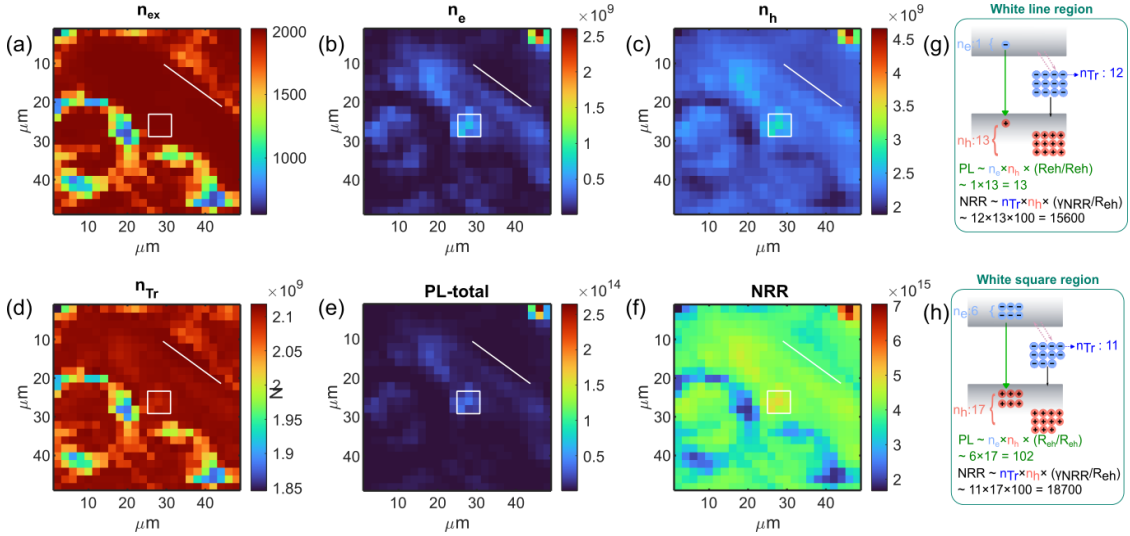

**Figure 5.** (a-f) Microscale mapping of the populations of excitons, electrons, holes, and filled traps ($n_{ex}$, $n_e$, $n_{Tr}$, $n_h$), along with the total PL emissions (PL-total) and nonradiative recombination loss (NRR) at 300 K. (g-h) Schematic representations of the carrier recombination mechanisms in the white line and white square regions, based on the carrier populations shown in (a–f).

Figure 5 provides a spatially resolved analysis of charge carrier recombination dynamics in MAPbBr$_3$ perovskite microcrystals. By mapping the populations of excitons, electrons, holes, and filled traps, along with total PL emissions and nonradiative recombination losses, this figure highlights the competition between radiative and nonradiative processes that dictate the overall photoluminescence quantum yield.

The exciton population ($n_{ex}$) shown in Figure 5a is significantly lower than the other carrier populations, consistent with the Saha-Langmuir equilibrium, where, at room temperature and under low excitation intensities, free charge carriers predominate over bound excitons in MAPbBr$_3$ perovskite films [15,16,18,47,55,56].

Since the exciton population (n$_{ex}$) is significantly smaller than the electron population (n$_e$) (Figure 5a-5b), the PL-ex is much less than the PL-eh (Figure S6), the total photoluminescence emission (PL-total) (Figure 5e) is predominantly attributed to electron-hole recombination (PL-eh). This relationship can be expressed as PL-total ≈ PL-eh ~ $R_{eh} n_e n_h$, while nonradiative recombination (Figure 5f) follows the trend NRR ~ $\gamma_{NRR} n_{Tr} n_h$. Given that the $R_{eh}$ and $\gamma_{NRR}$ are fixed parameters (from Table 1) and the hole population ($n_h$) is the shared population in both radiative and nonradiative recombination processes, their ratio simplifies to:

$$\frac{\text{PL--total}}{\text{NRR}} \sim \frac{R_{eh} n_e}{\gamma_{NRR} n_{Tr}} \sim \frac{R_{eh}}{\gamma_{NRR}} \frac{n_e}{n_{Tr}} = \frac{1}{100} \frac{n_e}{n_{Tr}}.$$

Therefore, the relative populations of electrons ($n_e$) and filled traps ($n_{Tr}$) generally influence the balance between PL emission and nonradiative depopulation losses.

To elucidate this balance, two regions—the white line region and the white square region—were selected for detailed investigation. These regions represent distinct recombination regimes based on their electron and filled trap populations. The average electron, hole, and filled trap populations for these regions are listed in Table S2. Figure 5g and 5h provide schematic visualizations of carrier recombination mechanisms in these regions, with simplified values presented in Table S3.

In the white line region (Figure 5g), the electron population (Figure 5b) is significantly lower than the filled trap population (Figure 5d), indicating a trap-predominated nonradiative recombination regime. From Table S3, we simplify the carrier distribution by considering 2 electrons in the conduction band, 13 holes in the valence band, and 11 filled traps in the trap states (Figure 5g). This results in a radiative PL-eh emission estimated as the product of electron and hole populations and normalized by the radiative recombination rate with a ratio $\frac{R_{eh}}{R_{eh}}$, calculated as 2 × 13 × 1 = 26. In contrast, the nonradiative recombination loss, determined by the filled trap population, hole population, and the nonradiative-to-radiative recombination rate ratio $\frac{\gamma_{NRR}}{R_{eh}}$, is estimated as 11 × 13 × 100 = 14300. The high NRR in this region confirms that trap-assisted nonradiative recombination significantly predominates over radiative recombination.

In the white square region (Figure 5h), the population of electron ($n_e$) increases a lot, while the population of filled traps ($n_{Tr}$) remain slightly more compared with the corresponding populations in the white line region. This increase in the electron population is primarily attributed to band-filling effects, which promote radiative recombination. Simplifying the carrier distribution, we assume 5 electrons in the conduction band, 17 holes in the valence band, and 12 filled traps in the trap states. This results in a calculated radiative PL-total emission of 5 × 17 × 1 = 85 and nonradiative recombination depopulation losses of 12 × 17 × 100 = 20400. While nonradiative recombination remains significant, the higher PL emission relative to the white line region suggests that radiative recombination plays a more prominent role due to increased electron population.

These results highlight the strong influence of trap states on charge carrier recombination processes. In regions with high trap densities, nonradiative recombination (NRR) predominates, leading to substantial PL quenching. Conversely, when trap densities are lower, band filling occurs, reducing the impact of defect-assisted nonradiative recombination and enhancing radiative efficiency. To further explore the relationship between defect properties and their effects on PL emission and nonradiative recombination losses, additional correlation analyses among carrier dynamics, radiative recombination, and nonradiative recombination losses were conducted across the entire film. These findings are systematically analyzed in Figure 6.

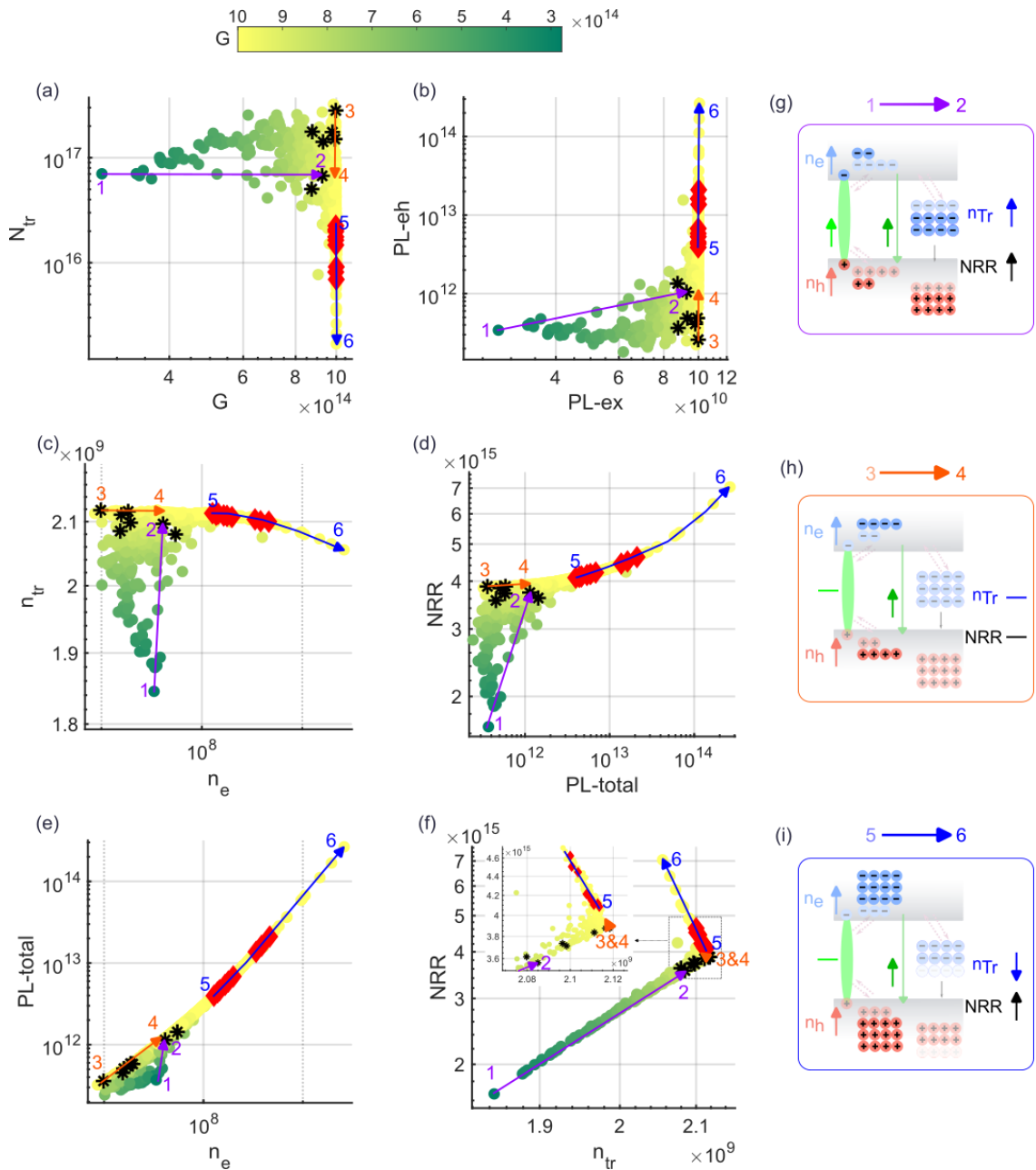

**Figure 6.** Correlation analysis between charge carrier populations, radiative recombination, and nonradiative recombination losses in MAPbBr$_3$ perovskite microcrystals. (a–f) Scatter plots of key physical parameters, color-coded according to increasing exciton generation rate ($G$) (from dark green for low excitation to bright yellow for high excitation): (a) exciton generation rate (G) versus total trap density ($N_{Tr}$), (b) PL-ex versus PL-eh, (c) electron population ($n_e$) versus filled trap population ($n_{Tr}$), (d) PL-total versus NRR, (e) electron population ($n_e$) versus PL-total, and (f) filled trap population ($n_{Tr}$) versus nonradiative NRR. (g–i) Schematic illustrations of three representative dynamic regimes highlighted in plots (a–f): (g) Trap-filling regime (trajectory 1→2). (h) Crossover regime (trajectory 3→4). (i) band-filling regime (trajectory 5→6).

Figure 6 provides detailed insights into the correlation between carrier dynamics, radiative recombination, and nonradiative losses in MAPbBr₃ perovskite microcrystalline films. The analysis is derived from simulated charge carrier recombination dynamics across an entire 576-pixel film map, representing the photoexcitation equilibrium conditions during IM2PM experiments at room temperature (see Section 2.3). Intrinsic local heterogeneity within the film gives rise to a wide distribution of generation rate of excitons (G) and total trap densities ($N_{Tr}$). The correlation plots (panels a–f) combined with schematic depictions (panels g–i) clearly shows three distinct recombination regimes.

Initially (trajectory 1→ 2), the exciton generation rate (G) increases while the total trap density ($N_{Tr}$) remains nearly constant (Figure 6a), leading to an increase in the overall charge carrier population, particularly in filled trap density ($n_{Tr}$), as shown in Figure 6c and 6f. Correspondingly, the nonradiative recombination (NRR) increases sharply (Figure 6d and 6f). Meanwhile, radiative PL contributions from exciton recombination (PL-ex) and free electron-hole recombination (PL-eh) remain relatively low (Figure 6b), highlighting the predominant role of nonradiative trap-assisted nonradiative recombination in this regime. The schematic in panel (g) illustrates the trap-filling scenario, where available traps are progressively filled, limiting radiative efficiency.

As G continues to increase, a subset of pixels enters a crossover regime (trajectory 3→4). These points, particularly concentrated around $N_{Tr}$ ~10¹⁷ cm⁻³, exhibit characteristics of both trap- and band-predominated recombination (see Figure 6f and inset). While trap states remain significantly filled, radiative recombination processes begin to emerge more clearly, evidenced by the rising PL-eh and PL-total (Figures 6b, 6d, and 6e). This transitional behavior indicates a competitive balance between trap-mediated nonradiative losses and radiative pathways. The schematic in panel (h) captures this intermediate state where both recombination channels are active.

During the trajectory from pixel 5 to pixel 6, the generation rate of excitons (G) remains constantly high while $N_{Tr}$ decreases (Figure 6a). Under these conditions, the total carrier population remains high, but the filled trap population ($n_{Tr}$) decreases significantly (Figure 6c), while the free electron density ($n_e$) increases sharply. This shift results in a marked enhancement of radiative recombination through PL-eh (Figures 6b, 6e). The increased $n_e$ also drives up the hole density ($n_h$), which—despite low filled trap density—leads to a modest increase in NRR due to NRR~$\gamma_{NRR} n_{Tr} n_h$ (Figure 6f). The schematic in panel (i) illustrates this radiatively favorable regime where band states are predominantly filled, and trap losses are minimized.

Altogether, the analysis of this single 576-pixel mapping highlights how local heterogeneity in exciton generation rate G and trap densities $N_{Tr}$ gives rise to spatially distinct recombination regimes within a nominally uniform film. These results demonstrate that performance-limiting processes such as trap-mediated recombination can vary dramatically at the microscale. By identifying and characterizing these regimes through spatial correlation analysis, this work provides a powerful framework for understanding and optimizing charge carrier dynamics in perovskite optoelectronics.

## 4 Conclusion

In this work, we developed and applied an optimized machine learning-assisted intensity-modulated two-photon photoluminescence microscopy (ml-IM2PM) framework to investigate charge carrier dynamics in MAPbBr₃ perovskite microcrystalline films. Balanced classification sampling was introduced to improve model robustness across a wide range of exciton generation rates, enabling accurate prediction of key parameters—including $G$, $N_{TR}$, and $E_a$ with spatial resolution across a 576-pixel film mapping.

Based on the predicted parameters, a system of ordinary differential equations (ODEs) was solved to simulate the charge carrier recombination processes under photoexcitation equilibrium condition at room temperature. The resulting spatial distributions of charge carrier populations, including excitons ($n_{ex}$), free electrons ($n_e$), holes ($n_h$), and filled traps ($n_{Tr}$)—as well as photoluminescence and nonradiative recombination losses, were quantitatively resolved across the film.

Through correlation analysis of the simulated data, three distinct recombination regimes were identified: (i) a trap-filling regime, where trap-assisted nonradiative losses is predominant; (ii) a crossover regime, characterized by simultaneous high G and high trap density $N_{Tr}$, where both trap- and band-mediated recombination processes coexist; and (iii) a band-filling regime, where high G and reduced trap density $N_{Tr}$ favor efficient radiative recombination and suppress nonradiative losses. Notably, we identified a critical threshold in trap density ($N_{Tr}$ ~$10^{17}$ cm$^{-3}$) marking the transition from the trap-predominated to the band-filling regime, emphasizing the decisive role of local defect states in controlling recombination processes

Together, these results demonstrate the effectiveness of ml-IM2PM as a high-resolution diagnostic tool for probing spatially heterogeneous carrier dynamics in perovskite films. The combination of machine learning, physical modeling, and spatial analysis provides a comprehensive platform for understanding and optimizing defect-related recombination pathways—offering valuable guidance for the design of high-efficiency, defect-tolerant perovskite optoelectronic devices.

## 5 Experimental Methods

**IM2PM setup**

The optical configuration used in this study is illustrated in **Figure S1**. The excitation source is a Ti:Sapphire mode-locked oscillator, producing broadband femtosecond laser pulses with a spectral range of 700–900 nm and a pulse duration of approximately 7 fs. To compensate for group velocity dispersion introduced by various optical elements, a pair of chirped mirrors is employed for dispersion management and pulse compression.

A Mach–Zehnder interferometer is used to introduce phase modulation into both arms of the beam path. Two acousto-optic modulators (AOMs), operating at 55 MHz and 54.95 MHz respectively, generate a modulation frequency difference of 50 kHz. This creates a periodically modulated average beam intensity, which is key to the intensity-modulated two-photon excitation scheme. After the third beam splitter, one portion of the beam is directed to an avalanche photodiode (APD) as a reference channel for monitoring laser intensity fluctuations in real time, while the other portion is directed into the sample path.

The sample is imaged using an inverted microscope equipped with a dichroic mirror that reflects wavelengths above 650 nm and transmits shorter wavelengths. A reflective objective (RO) with a numerical aperture (NA) of 0.65 is used to tightly focus the laser beam onto the sample surface. The two-photon-excited photoluminescence (PL) signal emitted from the sample is collected by an APD with a detection bandwidth of approximately 5 MHz.

To study the temperature dependence of the PL dynamics, the sample is mounted in a temperature-controlled stage (LTS420E-P, Linkam Scientific Instruments), allowing precise control over a range of discrete temperatures: 298, 273, 253, 233, 213, 193, 173, 153, 133, and 113 K. For simplicity, temperature values referenced throughout the manuscript are rounded to the nearest integer.

**Acknowledgments**

We acknowledge financial support from Swedish Energy Agency grant 50709-1, Swedish Research Council grant 2021-05207, KAW WISE/WASP grant 01-22, Olle Engkvist foundation grant 235-0422 and LU profile area Light and Materials. Collaboration within NanoLund is acknowledged. The authors thank K. J. Karki and P. Kumar for their valuable help with the experiment. The data handling was enabled by resources provided by LUNARC. ChatGPT was used to polish parts of the text of the article. We thank Junsheng Chen and Kaibo Zheng for the valuable discussions.

**Supporting Information**

S1. Optical setup.
S2 Charge carrier recombination parameters for IM2PM simulation in Figure 2
S3. Experimental raw data
S4 ml-IM2PM Workflow and Optimization
S5. Error estimation
S6. ml-IM2PM parameters prediction and ml-IM2PM assisted retrieved charge carrier recombination processes for white line and square regions
S7. Microscale mapping of the PL emission from excitons, electron-hole pairs, and the sum.

**References**

[1]  G. Hodes, *Science (1979)* **2013**, *342*, 317.

[2]  Best Research-Cell Efficiency Chart | Photovoltaic Research | NREL, https://www.nrel.gov/pv/cell-efficiency.html, accessed: Nov., 2023.

[3]  J. Kim, J. Heo, G. Park, S. Woo, C. Cho, *Nature* **2022**, *611*, 688.

[4]  X. Chang, J. Fang, Y. Fan, T. Luo, H. Su, Y. Zhang, J. Lu, L. Tsetseris, T. D. Anthopoulos, *Adv Mater* **2020**, *32*, 2001243.

[5]  S. D. Stranks, *ACS Energy Lett* **2017**, *2*, 1515.

[6]  J. Huang, Y. Yuan, Y. Shao, Y. Yan, *Nat. Rev. Mater.* **2017**, *2*, 1.

[7]  S. D. Stranks, V. M. Burlakov, T. Leijtens, J. M. Ball, A. Goriely, H. J. Snaith, *Phys Rev Appl* **2014**, *2*, 034007.

[8]  M. J. Trimpl, A. D. Wright, K. Schutt, L. R. V Buizza, Z. Wang, M. B. Johnston, H. J. Snaith, P. Müller-Buschbaum, L. M. Herz, *Adv Funct Mater* **2020**, *30*, 2004312.

[9]  J. Chen, K. Žídek, P. Chábera, D. Liu, P. Cheng, L. Nuuttila, M. J. Al-Marri, H. Lehtivuori, M. E. Messing, K. Han, K. Zheng, T. Pullerits, *J. Phys. Chem. Lett.* **2017**, *8*, 2316.

[10] K. Zheng, K. Žídek, M. Abdellah, M. E. Messing, M. J. Al-Marri, T. Pullerits, *J. Phys. Chem. C* **2016**, *120*, 3077.

[11] C. Stavrakas, A. A. Zhumekenov, R. Brenes, M. Abdi-Jalebi, V. Bulović, O. M. Bakr, E. S. Barnard, S. D. Stranks, *Energy Environ Sci* **2018**, *11*, 2846.

[12] D. W. DeQuilettes, S. M. Vorpahl, S. D. Stranks, H. Nagaoka, G. E. Eperon, M. E. Ziffer, H. J. Snaith, D. S. Ginger, *Science (1979)* **2015**, *348*, 683.

[13] Q. Shi, S. Ghosh, A. S. Sarkar, P. Kumar, Z. Wang, S. K. Pal, T. Pullerits, K. J. Karki, *J. Phys. Chem. C* **2018**, *122*, 3818.

[14] B. Yang, J. Chen, Q. Shi, Z. Wang, M. Gerhard, A. Dobrovolsky, I. G. Scheblykin, K. J. Karki, K. Han, T. Pullerits, *J. Phys. Chem. Lett.* **2018**, *9*, 5017.

[15] P. Kumar, Q. Shi, K. J. Karki, *J. Phys. Chem. C* **2019**, *123*, 13444.

[16] Q. Shi, P. Kumar, T. Pullerits, *ACS Phys. Chem. Au.* **2023**, *3*, 467.

[17] X. Du, L. Lüer, T. Heumueller, J. Wagner, C. Berger, T. Osterrieder, J. Wortmann, S. Langner, U. Vongsaysy, M. Bertrand, N. Li, T. Stubhan, J. Hauch, C. J. Brabec, *Joule* **2021**, *5*, 495.

[18] Q. Shi, T. Pullerits, *ACS Photonics* **2024**, *11*, 1093.

[19] Z. Zhang, J. Wang, Y. Zhang, J. Xu, R. Long, *J. Phys. Chem. Lett.* **2022**, *13*, 10734.

[20] K. T. Butler, D. W. Davies, H. Cartwright, O. Isayev, A. Walsh, *Nature 2018 559:7715* **2018**, *559*, 547.

[21] S. G. Motti, D. Meggiolaro, S. Martani, R. Sorrentino, A. J. Barker, F. De Angelis, A. Petrozza, S. G. Motti, S. Martani, R. Sorrentino, A. J. Barker, A. Petrozza, D. Meggiolaro, F. De Angelis, *Adv. Mater* **2019**, *31*, 1901183.

[22]   A. Mannodi-Kanakkithodi, J. S. Park, N. Jeon, D. H. Cao, D. J. Gosztola, A. B. F. Martinson, M. K. Y. Chan, *Chem. Mater.* **2019**, *31*, 3599.

[23]   J. M. Ball, A. Petrozza, *Nat. Energy* **2016**, *1*, 1.

[24]   S. Ghosh, S. K. Pal, K. J. Karki, T. Pullerits, *ACS Energy Lett* **2017**, *2*, 2133.

[25]   J. M. Azpiroz, E. Mosconi, J. Bisquert, F. De Angelis, *Energy Environ Sci* **2015**, *8*, 2118.

[26]   E. Mosconi, D. Meggiolaro, H. J. Snaith, S. D. Stranks, F. De Angelis, *Energy Environ Sci* **2016**, *9*, 3180.

[27]   Y. Tian, M. Peter, E. Unger, M. Abdellah, K. Zheng, T. Pullerits, A. Yartsev, V. Sundström, I. G. Scheblykin, *Phys. Chem. Chem. Phys.* **2015**, *17*, 24978.

[28]   M. D. McCluskey, M. D. McCluskey, E. E. Haller, E. E. Haller, *Dopants and Defects in Semiconductors* **2018**, DOI: 10.1201/B21986.

[29]   Q. Zhou, B. Wang, R. Meng, J. Zhou, S. Xie, X. Zhang, J. Wang, S. Yue, B. Qin, *Adv Funct Mater* **2020**, *30*, 2000550.

[30]   D. Meggiolaro, S. G. Motti, E. Mosconi, A. J. Barker, J. Ball, C. Andrea Riccardo Perini, F. Deschler, A. Petrozza, F. De Angelis, *Energy Environ Sci* **2018**, *11*, 702.

[31]   J. Dacuña, A. Salleo, *Phys Rev B Condens Matter* **2011**, *84*, DOI: 10.1103/PHYSREVB.84.195209.

[32]   G. Landi, Sergio Pagano, Heinz Christoph Neitzert, Costantino Mauro, Carlo Barone, *Energies (Basel)* **2023**, *16*, 1296.

[33]   D. Yang, W. Ming, H. Shi, L. Zhang, M. H. Du, *Chem. Mater.* **2016**, *28*, 4349.

[34]   N. Droseros, G. Longo, J. C. Brauer, M. Sessolo, H. J. Bolink, N. Banerji, *ACS Energy Lett* **2018**, *3*, 1458.

[35]   D. M. Niedzwiedzki, M. Kouhnavard, Y. Diao, J. M. D'Arcy, P. Biswas, *Phys. Chem. Chem. Phys.* **2021**, *23*, 13011.

[36]   Y. Yang, M. Yang, Z. Li, R. Crisp, K. Zhu, M. C. Beard, *J. Phys. Chem. Lett.* **2015**, *6*, 4688.

[37]   Tomas Leijtens, G. E. Eperon, A. J. Barker, Giulia Grancini, Wei Zhang, J. M. Ball, A. R. Srimath Kandada, H. J. Snaith, Annamaria Petrozza, *Energy Environ Sci* **2016**, *9*, 3472.

[38]   G.-J. A. H. Wetzelaer, M. Scheepers, A. M. Sempere, C. Momblona, J. Ávila, H. J. Bolink, *Adv Mater* **2015**, *27*, 1837.

[39]   Kiligaridis A, Frantsuzov PA, Yangui A, Seth S, Li J, An Q, Vaynzof Y, Scheblykin IG, *Nat. Commun* **2021**, *12*, DOI: 10.1038/s41467-021-23275-w.

[40]   S. Feldmann, S. Macpherson, S. P. Senanayak, M. Abdi-Jalebi, J. P. H. Rivett, G. Nan, G. D. Tainter, T. A. S. Doherty, K. Frohna, E. Ringe, R. H. Friend, H. Sirringhaus, M. Saliba, D. Beljonne, S. D. Stranks, F. Deschler, *Nat. Photonics* **2019**, *14*, 123.

[41]   Z. Andaji-Garmaroudi, M. Anaya, A. J. Pearson, S. D. Stranks, *Adv Energy Mater* **2020**, *10*, 1903109.

[42]   R. J. Elliott, *Phys. Rev.* **1957**, *108*, 1384.

[43] C. L. Davies, M. R. Filip, J. B. Patel, T. W. Crothers, C. Verdi, A. D. Wright, R. L. Milot, F. Giustino, M. B. Johnston, L. M. Herz, *Nat Commun* **2018**, *9*, 293.

[44] C. S. Ponseca, T. J. Savenije, M. Abdellah, K. Zheng, A. Yartsev, T. Pascher, T. Harlang, P. Chabera, T. Pullerits, A. Stepanov, J. P. Wolf, V. Sundström, *J Am Chem Soc* **2014**, *136*, 5189.

[45] K. Zheng, Q. Zhu, M. Abdellah, M. E. Messing, W. Zhang, A. Generalov, Y. Niu, L. Ribaud, S. E. Canton, T. Pullerits, *J. Phys. Chem. Lett.* **2015**, *6*, 2969.

[46] V. D'Innocenzo, G. Grancini, M. J. P. Alcocer, A. R. S. Kandada, S. D. Stranks, M. M. Lee, G. Lanzani, H. J. Snaith, A. Petrozza, *Nat Commun* **2014**, *5*, 1.

[47] Q. Shi, S. Ghosh, P. Kumar, L. C. Folkers, S. K. Pal, T. Pullerits, K. J. Karki, *J. Phys. Chem. C* **2018**, *122*, 21817.

[48] V. Al Osipov, X. Shang, T. Hansen, T. Pullerits, K. J. Karki, *Phys. Rev. A* **2016**, *94*, 053845.

[49] S. Liu, H. Hua, *Opt. Express* **2011**, *19*, 353.

[50] Field Guide to Microscopy | (2010) | Tkaczyk | Publications | SPIE, https://spie.org/publications/book/798239, accessed: Nov., 2024.

[51] V. M. Le Corre, E. A. Duijnstee, O. El Tambouli, J. M. Ball, H. J. Snaith, J. Lim, L. J. A. Koster, *ACS Energy Lett* **2021**, *6*, 1087.

[52] B. Wenger, P. K. Nayak, X. Wen, S. V Kesava, N. K. Noel, H. J. Snaith, *Nat Commun* **2017**, *8*, 590.

[53] G. Han, T. M. Koh, S. S. Lim, T. W. Goh, X. Guo, S. W. Leow, R. Begum, T. C. Sum, N. Mathews, S. Mhaisalkar, *ACS Appl Mater Interfaces* **2017**, *9*, 21292.

[54] A. Nur'aini, S. Lee, I. Oh, *JECST* **2021**, *13*, 71.

[55] M. J. Dresser, *J Appl Phys* **1968**, *39*, 338.

[56] S. Ghosh, Q. Shi, B. Pradhan, P. Kumar, Z. Wang, S. Acharya, S. K. Pal, T. Pullerits, K. J. Karki, *J. Phys. Chem. Lett.* **2018**, *9*, 4245.

# Supporting Information

# High Resolution AI-Enhanced Functional Imaging Reveals Heterogeneity of Trap States and Charge Carrier Recombination Pathways in Perovskite

*Qi Shi[1,2]\*, and Tönu Pullerits [1,2]\**

[1] The Division of Chemical Physics, Department of Chemistry, Lund University, Kemicentrum Naturvetarevägen 16, 223 62 Lund, Sweden

[2] NanoLund, Department of Chemistry, Lund University, Professorsgatan 1, 223 63 Lund, Sweden

AUTHOR INFORMATION

**Corresponding Author**

\*qi.shi@chemphys.lu.se

\*Tonu.Pullerits@chemphys.lu.se

## S1. Optical setup

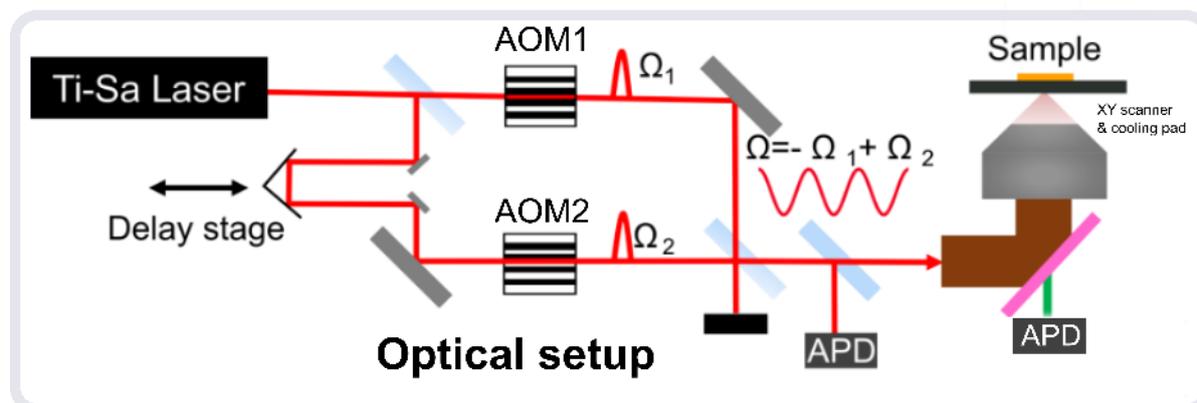

**Figure S1** optical setup.

**Figure S1** illustrates the optical configuration utilized in this investigation. The light source comprises a Ti-Sapphire laser-based mode-locked oscillator, generating broadband laser pulses with a spectral range spanning from 700 nm to 900 nm and a pulse duration of approximately 7 fs. To compensate for group velocity dispersion introduced by optical components throughout the setup, a pair of chirped mirror pairs is employed to mitigate this effect.

Within a Mach-Zehnder interferometer, a time-dependent phase change is introduced into each arm using two acousto-optic modulators (AOM) to modulate the average laser beam intensity. Operating at radio wave frequencies of 55 MHz and 54.95 MHz, with a frequency difference of 50 KHz, the AOMs facilitate this modulation. Following the third beam splitter, one portion is detected by an avalanche photodiode (APD) and serves as a reference, enabling the monitoring of laser intensity fluctuations. Simultaneously, the other portion is directed to an inverted microscope.

The microscope features a dichroic mirror reflecting light wavelengths longer than 650 nm and transmitting shorter wavelengths. A reflective objective (RO) with a numerical aperture of 0.65 precisely focuses the laser beam onto the sample. Two-photon-induced photoluminescence (PL) intensity is detected by an avalanche photodiode (APD) with a bandwidth of approximately 5 MHz.

To enable controlled temperature variations in the sample, a temperature-controlled type (Linkam Scientific Instruments, LTS420E-P) is employed, allowing adjustments at temperatures of 298, 273, 253, 233, 213, 193, 173, 153, 133, and 113 K. Please note that temperature values in this work are rounded for simplicity.

SEM image of MAPbBr$_3$ perovskite microcrystalline film

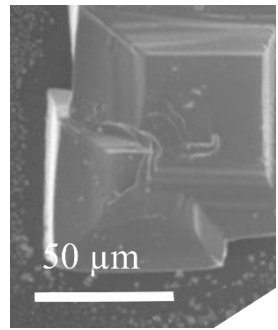

**Figure S2** The SEM image of the MAPbBr$_3$ perovskite microcrystalline film.

S2 Charge carrier recombination parameters for IM2PM simulation in Figure 2

**Table S1** Summary of charge carrier recombination parameters for simulation of the charge carrier recombination dynamics based on IM2PM in Figure 2 at room temperature. *RT* refers to room temperature.

| Recombination Parameters | Values |
|---|---|
| Exciton formation rate $R_f$ | $10^{-12}\ cm^3 s^{-1}$ |
| Exciton dissociation rate $R_d(RT)$ | $5 \times 10^{11}\ s^{-1}$ |
| Trapping rate $\gamma_{trap}$ | $8 \times 10^{-9}\ cm^3 s^{-1}$ |
| Detrapping rate $\gamma_{detrap}(RT)$ | $10^7\ s^{-1}$ |
| Electron hole radiative recombination rate $R_{eh}$ | $5 \times 10^{-11}\ cm^3 s^{-1}$ |
| Exciton radiative recombination rate $R_{ex}$ | $5 \times 10^7\ s^{-1}$ |
| Shockley-Read-Hall (SRH) nonradiative recombination rate $\gamma_{NRR}(RT)$ | $5 \times 10^{-9}\ cm^3 s^{-1}$ |
| Initial trap density ($N_{TR}$) | $2.0 \times 10^{17}\ cm^3$ |
| Generation of exciton (G) | $1 \times 10^{15}\ cm^3$ |
| Total trap concentration at RT ($N_{tr}$) | $8.0 \times 10^{14}\ cm^3$ |
| Trap energy barrier ($E_a$) | 67 meV |

**S3. Experimental raw data**

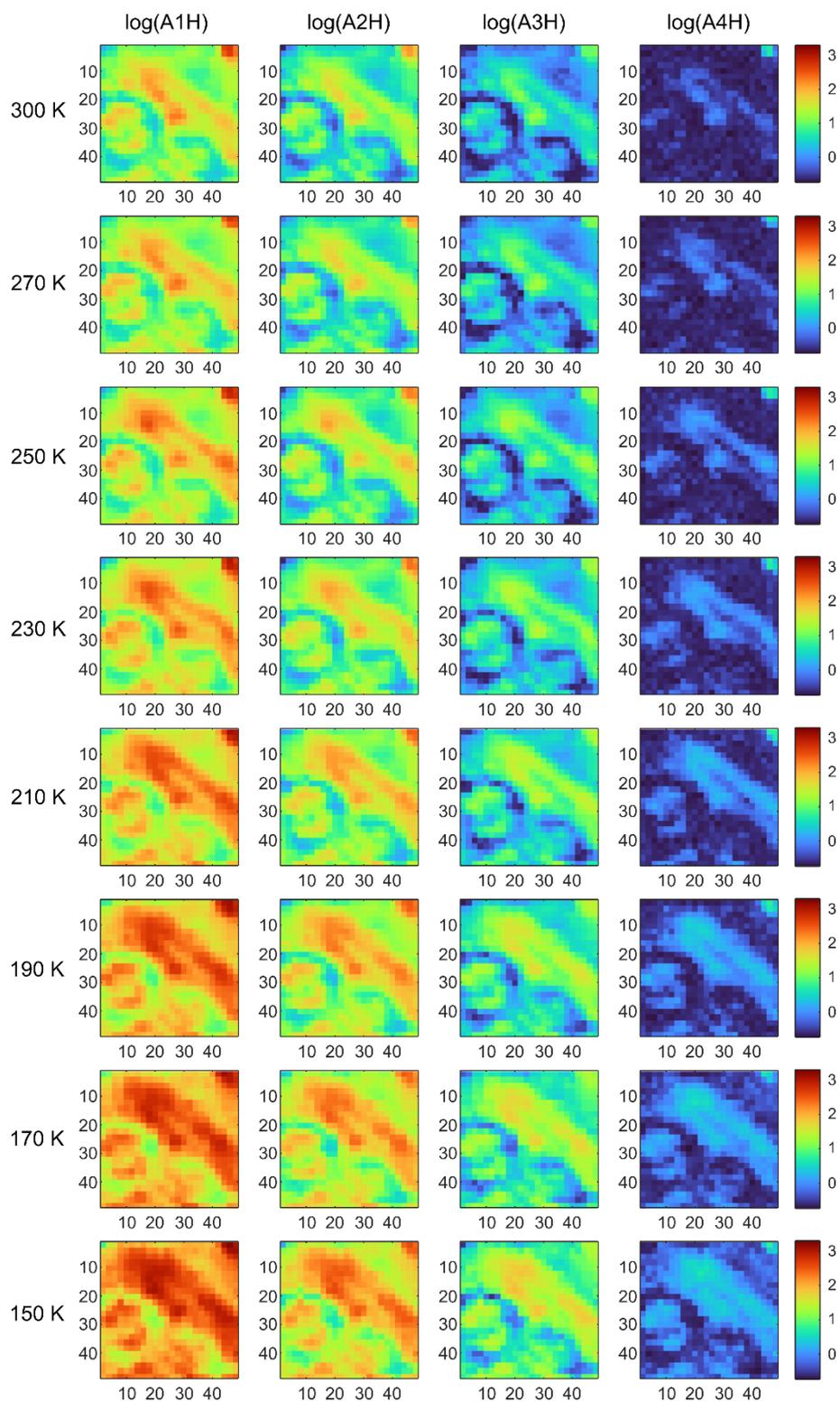

**Figure S3** Experimental raw data A1H A2H A3H A4H images at 8 temperatures.

**S4 ml-IM2PM Workflow and Optimization**

Table S2 Comparison of model performance across three G sampling strategies in ml-IM2PM cross-validation.

| | Chain-ET | G linear | G linear 2$^{nd}$ value | G log |
|---|---|---|---|---|
| R2 coefficient | G | 0.99 | 0.99 | 0.99 |
|  | NTR | 0.97 | 0.97 | 0.96 |
|  | Ea | 0.87 | 0.84 | 0.82 |
| PCC coefficient Compare with Experimental data | Mean_8T | 0.928 | 0.939 | 0.950 |
|  | RT | 0.967 | 0.960 | 0.974 |

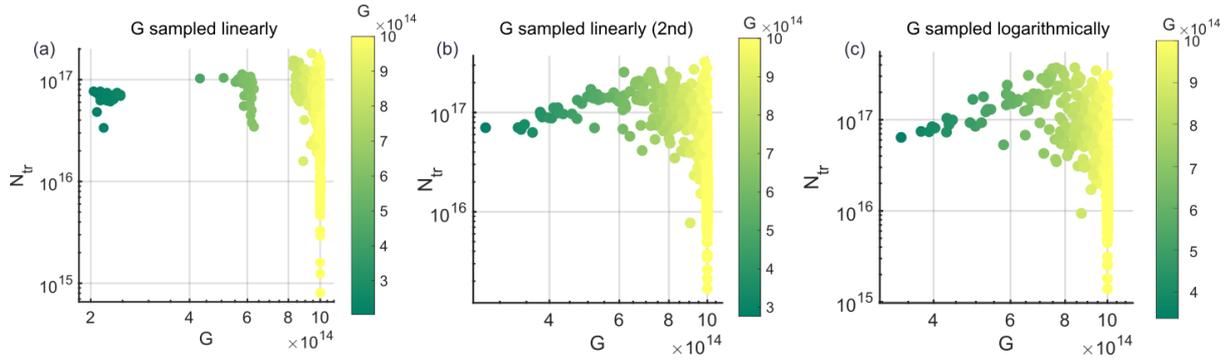

Figure S4 Correlation between exciton generation rate (G) and effective trap density ($N_{Tr}$) at room temperature reveals sampling imbalance in initial parameter selection.

This section provides a detailed description of the machine learning-assisted IM2PM (ml-IM2PM) workflow used to extract spatially resolved recombination parameters in MAPbBr$_3$ perovskite microcrystalline films.

The method involves simulating a large dataset of temperature dependent PL Fourier spectra (AiH) with varying three key parameters: exciton generation rate (G), trap energy barrier ($E_a$), and initial trap density ($N_{TR}$). These parameters were systematically sampled across physically relevant ranges, and the resulting simulated temperature dependent PL (AiH) spectra were used to train a regression model based on a Chain - Extra Trees ensemble algorithm. The trained ml-IM2PM model was then applied to temperature dependent experimental PL (AiH) maps to predict the spatial distributions of G, $E_a$, and $N_{TR}$ across the perovskite microcrystalline film. The trained model demonstrates strong predictive performance on unseen data, with R² values of 0.87, 0.99, and 0.97 for G, $E_a$, and $N_{TR}$, respectively.

In the second step, the trained model is applied to temperature dependent experimental PL (AiH) maps to predict spatial distributions of recombination parameters of the ODEs model across the perovskite films. The accuracy of the predicted parameters is validated by reconstructing temperature-dependent PL (AiH) spectra from the ml-IM2PM predictions and comparing them with experimental PL (AiH) measurements. The high average Pearson correlation coefficient (PCC) of 0.928 across eight temperatures—and 0.967 at room temperature—demonstrates the model's robustness and accuracy in reproducing experimental photophysical behaviour.

Both revised strategies underwent the same three-step ml-IM2PM workflow. Results presented in Table S3 show that the third strategy—logarithmic G sampling—yields the highest fidelity in reconstructing experimental PL (AiH) spectra. Moreover, the correlation plots confirm that both Strategies 2 and 3 eliminate the imbalance seen in the initial sampling. Based on these improvements in prediction accuracy and parameter distribution uniformity, the logarithmic **G** sampling approach (Strategy 3) was selected for all subsequent analyses.

## S5. Error estimation

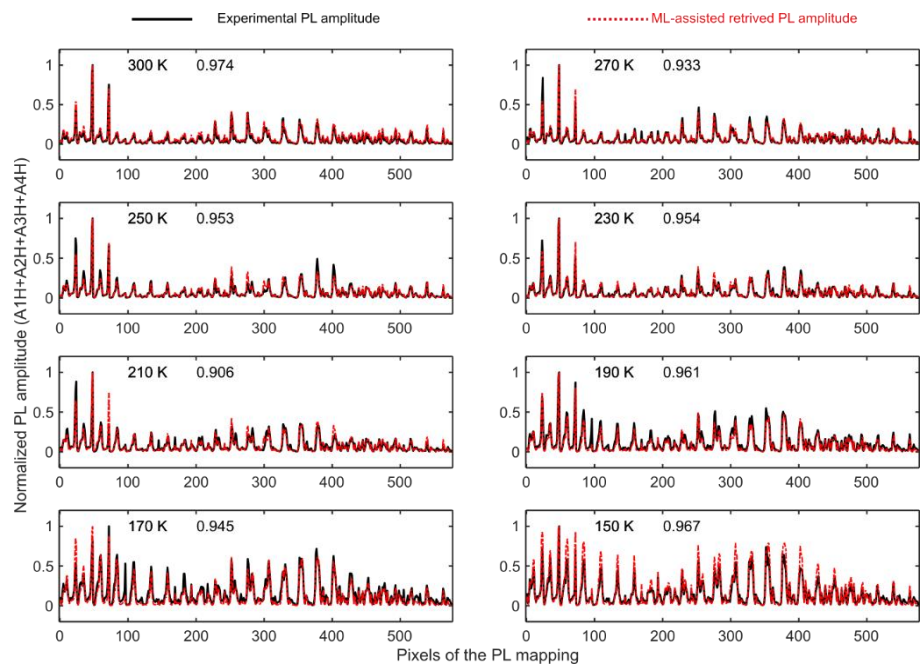

**Figure S4** The temperature dependent normalized experimental PL data (A1H+A2H+A3H+A4H), is represented by the black line. The temperature dependent PL data retrieved with ml-IM2PM assistance (A1H+A2H+A3H+A4H) is depicted by the red dotted line. The x axis is the pixels of the PL mapping (576 pixels (24×24) in the investigated area 48×48 µm$^2$ with a 2 µm scanning step). The y axis is the normalized total modulated PL emission amplitude. PCC: Pearson correlation coefficient.

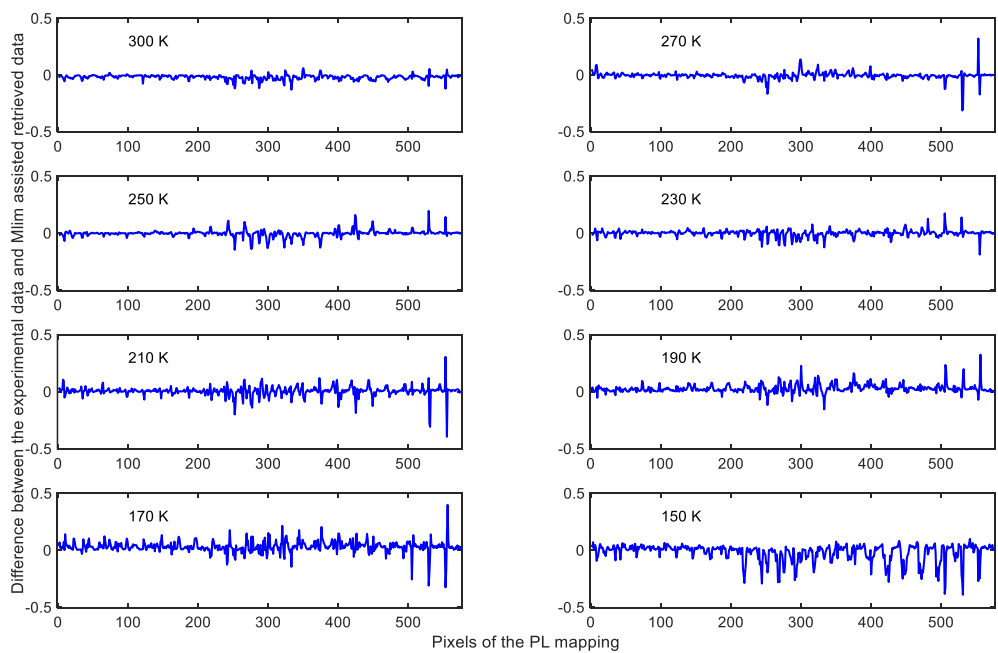

**Figure S5** the temperature dependent difference between the experimental PL emission (A1H+A2H+A3H+A4H) and the ml-IM2PM assisted retrieved PL emission (A1H+A2H+A3H+A4H). The x axis is the pixels of the PL mapping (576 pixels (24×24) in the investigated area 48×48 μm² with a 2 μm scanning step).

## S6. ml-IM2PM parameters prediction and ml-IM2PM assisted retrieved charge carrier recombination processes for white line and square regions

**Table S3** The ml-IM2PM prediction of generation of exciton (G), trap concentration (Ntr) at room temperature, and the ml-IM2PM assisted retrieved modulated exciton population, electron population, hole population, filled trap concentration, PL emission from excitons, PL emission from electron hole pairs, total PL emissions, and nonradiative depopulations losses for the averaged white line region, white square region in Figure 6, respectively.

| RT (300 k) Population ($cm^3$) | White line region | White square region | 1 | 2 | 3 | 4 | 5 | 6 |
|---|---|---|---|---|---|---|---|---|
| Generation of exciton (G) $\times 10^{15}$ | 1 | 1 | 0.3 | 0.9 | 1 | 1 | 1 | 1 |
| Trap concentration (Ntr) $\times 10^{16}$ | 13.2 | 0.7 | 7.0 | 6.6 | 17.0 | 6.2 | 7.9 | 1.7 |
| Exciton population $\times 10^3$ | 2.0 | 2.0 | 0.6 | 1.8 | 2.0 | 2.0 | 2.0 | 2.0 |
| Electron population $\times 10^8$ | 0.3 | 4.8 | 0.3 | 0.41 | 0.2 | 0.5 | 0.36 | 26.0 |
| Hole population $\times 10^8$ | 21.4 | 25.7 | 18.8 | 21.3 | 21.3 | 21.6 | 21.5 | 46.6 |
| Filled trap concentration $\times 10^8$ | 21.1 | 21.0 | 18.5 | 20.8 | 21.1 | 21.1 | 21.1 | 20.6 |
| PL-ex $\times 10^{11}$ | 1.0 | 1.0 | 0.3 | 0.9 | 1.0 | 1.0 | 1.0 | 1.0 |
| PL-eh $\times 10^{11}$ | 6.8 | 221.7 | 3.4 | 10.2 | 4.3 | 12.3 | 9.5 | 2646.1 |
| PL-total $\times 10^{11}$ | 7.8 | 222.7 | 3.7 | 11.1 | 5.3 | 13.3 | 10.5 | 2647.1 |
| Nonradiative recombination loss $\times 10^{14}$ | 39.0 | 46.0 | 16.7 | 36.2 | 38.5 | 39.4 | 39.0 | 70.5 |

**Table S4** Simplified value table.

| RT (300 k) | White line region | White square region |
|---|---|---|
| Electron population | 1 | 6 |
| Hole population | 13 | 17 |
| Filled trap concentration | 12 | 11 |

## S7. Microscale mapping of the PL emission from excitons, electron-hole pairs, and the sum.

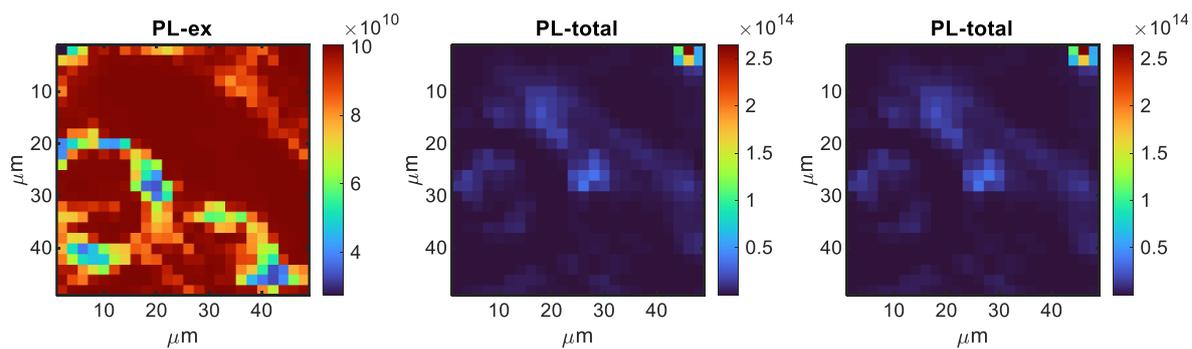

**Figure S6** Microscale mapping of the PL emission from excitons (PL-ex), PL emission from electron hole pairs (PL-eh) and total PL emission (PL-total).